%% file: main.tex
\input{components/style}

\begin{document}
% \vspace*{0.2in}

\input{components/head}
\input{sections/abstract}

% \linenumbers

\input{sections/introduction}
\input{sections/framework}

\input{sections/apl}

\input{sections/conclusion}
\input{sections/availability}
\input{sections/acknowledgement}
% \nolinenumbers

% \input{sections/reference}
\bibliography{reference}

\end{document}

%% file: components/style.tex
\documentclass[10pt,letterpaper]{article}
\usepackage[margin=1in]{geometry}

\usepackage{amsmath,amssymb}

\usepackage{changepage}

\usepackage{textcomp,marvosym}

\usepackage{cite}

\usepackage{nameref,hyperref}

\usepackage[nopatch=eqnum]{microtype}
\DisableLigatures[f]{encoding = *, family = * }

\usepackage[table]{xcolor}

\usepackage{array}

\usepackage[aboveskip=1pt,labelfont=bf,labelsep=period,justification=raggedright,singlelinecheck=off]{caption}

\makeatletter
\renewcommand{\@biblabel}[1]{\quad#1.}
\makeatother

\usepackage{lastpage,fancyhdr,graphicx}
\usepackage{epstopdf}

%% file: components/head.tex
\begin{flushleft}
{\Large
\textbf\newline{APLSuite: An Integrated Suite for CD4+ T Cell Epitope Prediction via Antigen Processing Likelihood} % Please use "sentence case" for title and headings (capitalize only the first word in a title (or heading), the first word in a subtitle (or subheading), and any proper nouns).
}
\newline
\\
Jiarui Li\textsuperscript{1},
Marco K. Carbullido\textsuperscript{1},
Jai Bansal\textsuperscript{2},
Samuel J. Landry\textsuperscript{3},
Ramgopal R. Mettu\textsuperscript{1,*},
\\
\bigskip
\textbf{1} Department of Computer Science, Tulane University, New Orleans, LA, United States
\\
\textbf{2} Isidore Newman School, New Orleans, LA, United States
\\
\textbf{3} Department of Biochemistry and Molecular Biology, Tulane University School of Medicine, New Orleans, LA, United States
\\

% Insert additional author notes using the symbols described below. Insert symbol callouts after author names as necessary.
% 
% Remove or comment out the author notes below if they aren't used.
%
% Primary Equal Contribution Note
% \Yinyang These authors contributed equally to this work.

% % Additional Equal Contribution Note
% % Also use this double-dagger symbol for special authorship notes, such as senior authorship.
% \ddag These authors also contributed equally to this work.

% % Current address notes
% \textcurrency Current Address: Dept/Program/Center, Institution Name, City, State, Country % change symbol to "\textcurrency a" if more than one current address note
% % \textcurrency b Insert second current address 
% % \textcurrency c Insert third current address

% % Deceased author note
% \dag Deceased

% % Group/Consortium Author Note
% \textpilcrow Membership list can be found in the Acknowledgments section.

% Use the asterisk to denote corresponding authorship and provide email address in note below.
* rmettu@tulane.edu \\
\bigskip
Project Page: \textcolor{magenta}{\url{https://tulane-mettu-landry-lab.github.io/blogs/APLSuite/}}

\end{flushleft}

%% file: sections/abstract.tex
\section*{Abstract}
Computational epitope prediction is a critical tool for exploring and understanding CD4+ T cell-mediated immune responses, a key aspect of adaptive immunity. While existing computational methods primarily focus on supervised learning approaches, they often overlook the essential role of antigen processing in determining binding specificity. To address this limitation, our group developed Antigen Processing Likelihood (APL), an algorithm that integrates crystallographic B-factor, solvent accessible surface area (SASA), hydrogen exchange protection factors (COREX), and sequence entropy. 

In this paper we introduce APLSuite, a comprehensive and lightweight software suite designed to streamline APL-based epitope prediction. APLSuite integrates distributed RESTful API services, a Python client for data aggregation and processing, a data science tool for efficient epitope computation, and a user-friendly graphical user interface for non-coding users. It provides a seamless and efficient pipeline for APL calculation and epitope prediction that can be finished in minutes with GPU-acceleration, which has not been implemented by existed tools. This flexible and extensible software suite is deployable on desktop and cloud environments, offering both guided and customizable workflows to meet diverse research needs in immunology research and immunotherapy development.

% Please keep the Author Summary between 150 and 200 words
% Use first person. PLOS ONE authors please skip this step. 
% Author Summary not valid for PLOS ONE submissions.   
% \section*{Author summary}
% Lorem ipsum dolor sit amet, consectetur adipiscing elit. Curabitur eget porta erat. Morbi consectetur est vel gravida pretium. Suspendisse ut dui eu ante cursus gravida non sed sem. Nullam sapien tellus, commodo id velit id, eleifend volutpat quam. Phasellus mauris velit, dapibus finibus elementum vel, pulvinar non tellus. Nunc pellentesque pretium diam, quis maximus dolor faucibus id. Nunc convallis sodales ante, ut ullamcorper est egestas vitae. Nam sit amet enim ultrices, ultrices elit pulvinar, volutpat risus.

%% file: sections/introduction.tex
\section*{Introduction}
CD4+ T cells are critical to adaptive immunity, playing a central role in mediating immune responses across diverse biological contexts~\cite{kumar2018human}. The computational prediction of antigenic peptides that are presented and bound to T cells has been an area of research for decades, which highly contributes to immunology research and immunotherapy development. While current approaches primarily utilize supervised learning techniques (e.g., NetMHC-II ~\cite{nilsson2023accurate,nilsson2023machine} provided by IEDB~\cite{vita2019immune,iedb.org}) to predict peptide-MHC-II binding, these methods typically do not account for the role of antigen processing.

To address this limitation, our group developed the Antigen Processing Likelihood (APL) algorithm ~\cite{mettu2016cd4+,charles2022cd4+,landry2023structural}, which models conformation stability using crystallographic B-factor (or AlphaFold pLDDT), solvent accessible surface area (SASA), correlations with hydrogen exchange protection factors (COREX) ~\cite{hilser1996structure}, and sequence entropy. Despite its robust theoretical framework, APL relies on multiple computational components, which currently lack integration into a single tool. As a result, researchers must compute each factor manually using separate tools and then combine the results using the APL algorithm. 

For SASA calculations, FreeSASA ~\cite{mitternacht2016freesasa} is the commonly used Python package. COREX is available through a web service ~\cite{vertrees2005corex}, but it is a single-processor, CPU-based algorithm, consuming hours to days for a single calculation. To overcome this bottleneck, our group developed both CPU-parallelized and GPU-parallelized versions of COREX ~\cite{bhattacharya2023parallel,jiarui2024gpu,jiarui2024gpumcmc}, which reduce computation times from hours or days to under 180 seconds. Sequence entropy calculations depend on BLAST ~\cite{altschul1990basic,camacho2009blast+} and Clustal Omega ~\cite{edgar2006multiple,sievers2011fast,sievers2018clustal} for sequence alignment, which require additional computational resources such as GPUs, CPUs, and access to large-scale databases.

These tools are provided by different sources, operate in varying formats, and demand diverse computational resources. To address these challenges, we present an integrated software suite that we call \textbf{APLSuite} (Fig~\ref{fig:framework}) that can run the entire APL pipeline seamless within minutes. It consists of a distributed RESTful API framework, a Python client for distributed API access, a data science tool, and a web application graphical user interface for non-coding users. The suite is designed to be flexible, extensible, and lightweight, enabling deployment on both desktop systems and cloud services. For users, the web front-end provides a step-by-step guided mode for ease of use, as well as a highly customizable mode for advanced applications. Leveraging the framework, we provided several APL and APL associated with MHC prediction automated pipelines, which accepting PDB file, PDB ID, and AlphaFold predicted mmCIF file as inputs separately. 

%% file: sections/framework.tex
\begin{figure}[t]%
\centering
\includegraphics[width=\linewidth]{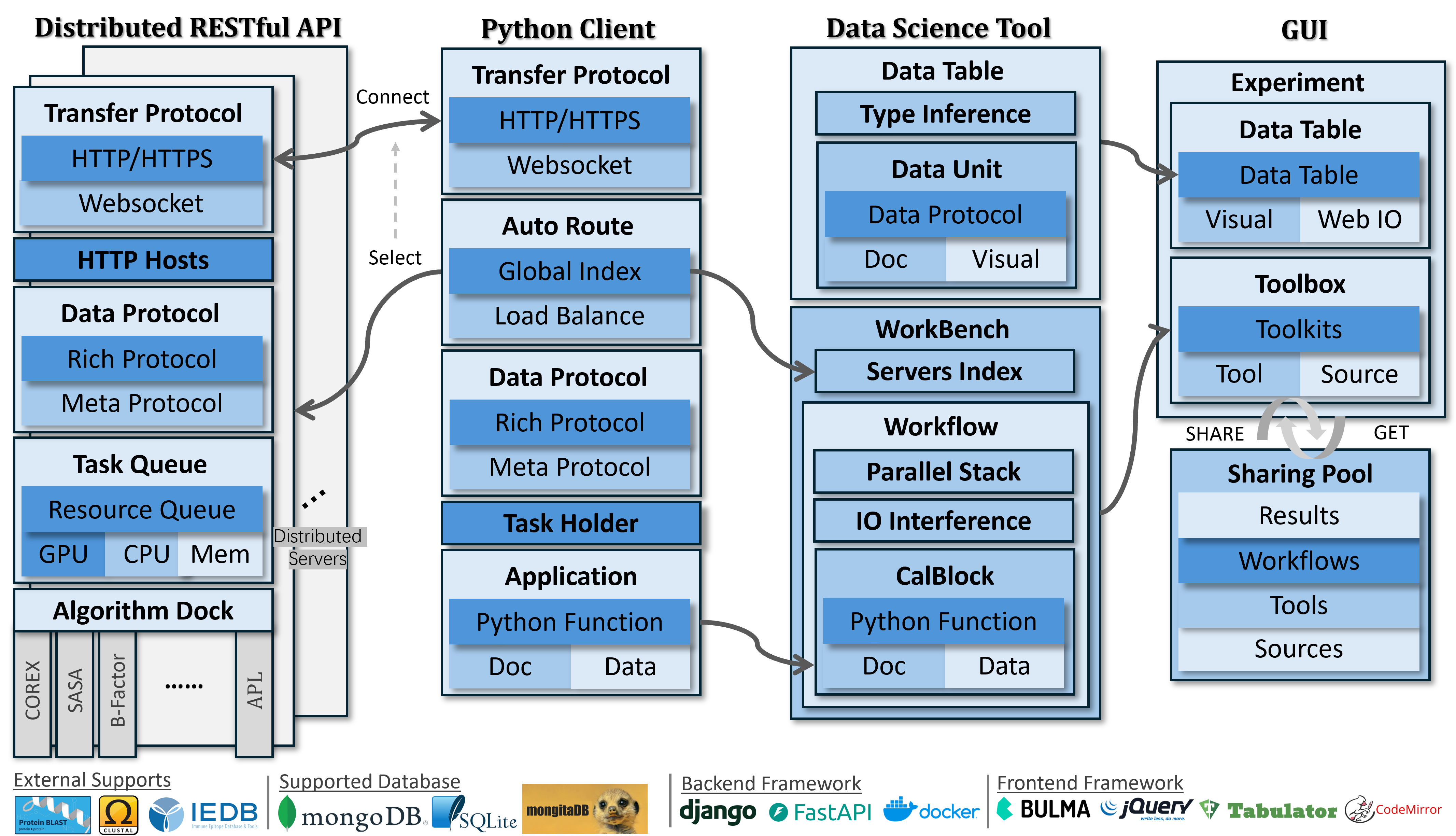}

\caption{\textbf{APLSuite Framework Overview: }The APLSuite framework including Distributed RESTful API (DRAF), Python client, data science tool, and graphical user interface (GUI). The APL components and itself are developed as API endpoints, remote Python function, and web UI based tool using this framework. It is network service based on Django, FastAPI, and Docker. The storage engine supporting SQLite, MongoDB, and MongitaDB. The computational services are also supported by BLAST~\cite{altschul1990basic,camacho2009blast+}, Clustal Omega ~\cite{edgar2006multiple,sievers2011fast,sievers2018clustal}, and IEDB~\cite{vita2019immune,iedb.org}.}
\label{fig:framework}
\end{figure}

\section*{Design and Implementation}
This application framework consists of four components including: Graphical User Interface (GUI) for non-coding users, Distributed RESTful API Framework (DRAF), Python client,  and Data Science Tool (DST) to support the GUI and coding users.

\subsection*{Graphical User Interface}

To accommodate non-coding users and simplify the utilization of APL computations, we have developed a Graphical User Interface (GUI) supported by the DRAF and DST frameworks, built on Django ~\cite{Django}. %This WGUI incorporates all functionalities of DRAF and DST, offering various operational modes tailored for both novice and experienced users. 
The GUI consists of five primary views: dashboard view, project view, tool view, quick start view, and resource management view. Each computation algorithm or chained algorithm is represented as a tool, while each computation case is managed as a project. Users can manage tools and projects centrally through the dashboard view.

\begin{figure}[ht]%
\centering
\includegraphics[width=\linewidth]{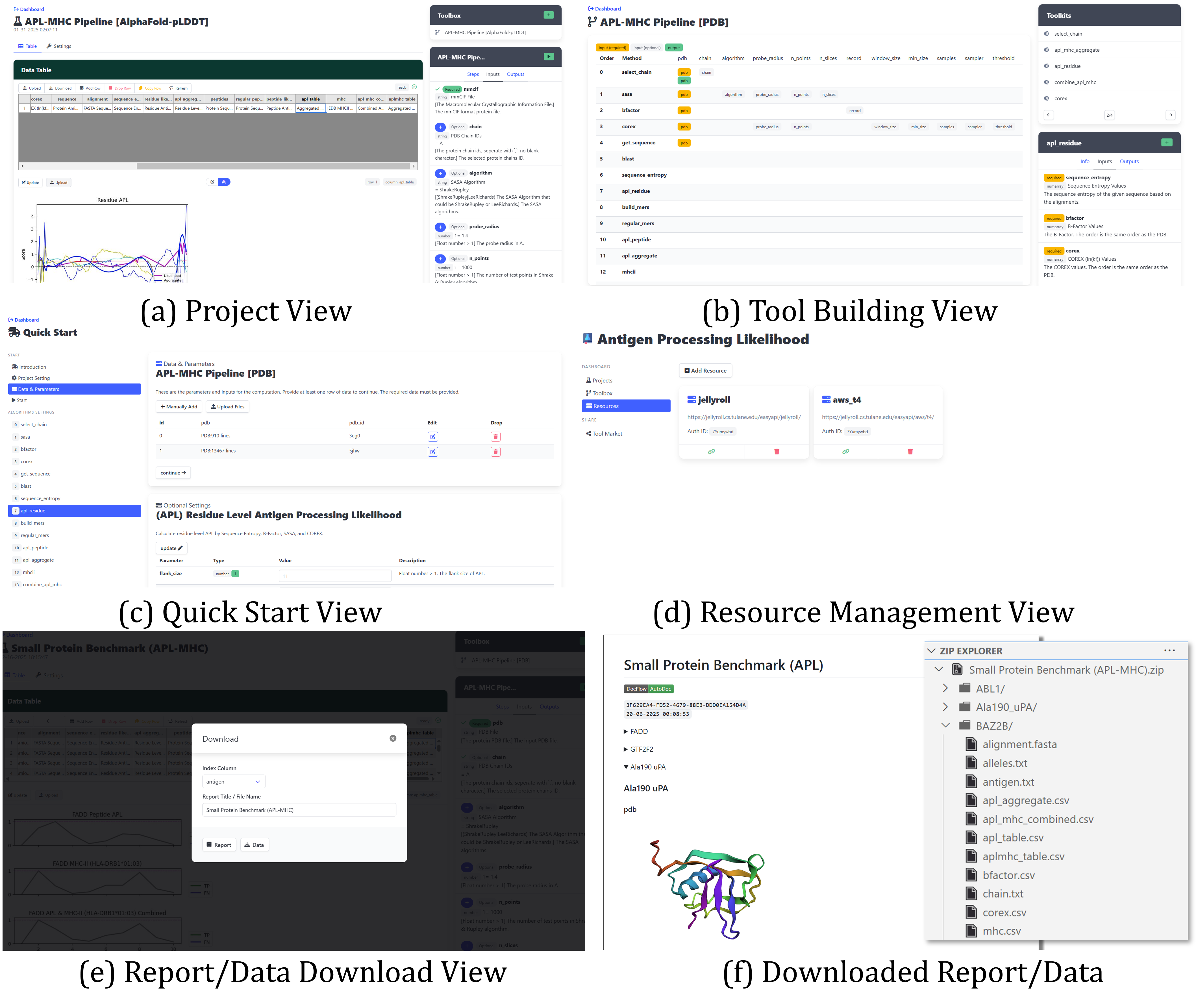}

\caption{\textbf{Graphical User Interface (GUI):} The views of GUI including project view, tool view, quick start view, and resource management view. Which enables both step-by-step guided running and highly flexible workbench. It also supports to download the results as a report or a compressed file of CSV tables.}
\label{fig:webui}
\end{figure}

The project view (Fig~\ref{fig:framework}.(a)) is designed for managing and executing specific computational tasks (projects). It includes four main components: data table, data visualization, toolbox, and job list.
The editable data table stores computational inputs (e.g., PDB files, PDB IDs), parameters (e.g., COREX thresholds, APL weights), and outputs. Users can edit the table directly by modifying values or uploading files to customize inputs or parameters. For any selected cell in the data table, the corresponding visualizations generated by DST are displayed, offering immediate feedback on the data.
The toolbox lists all tools available for the current project (e.g., COREX, SASA, APL). Users can explore each tool’s algorithm combinations, descriptions, inputs, and outputs. If parameter customization is required, users can add specific parameters to the data table by clicking the ``Add'' button. Once configured, users can execute the tool by clicking ``Run'', which queues the task in the job list.
The job list tracks all tasks, displaying their status (e.g., in-queue, running, error) and providing specific error messages when applicable.
This view offers researchers a fully customizable and flexible interface for running individual components or the entire APL pipeline with their preferred parameters.

For users seeking a simplified experience, the quick start view (Fig~\ref{fig:webui}.(c)) provides a step-by-step guided interface. Users are prompted to input or upload required files (e.g., PDB files for the APL pipeline). Computations can then be initiated directly. While parameter customization is supported, it is optional and can be skipped, making this view ideal for users who prioritize simplicity over customization.

The tool view (Fig~\ref{fig:webui}.(b)) allows users to build or customize tools and tool chains. Users can assemble their own tools by combining existing blocks, ensuring that the inputs and outputs of adjacent blocks share the same name. The system automatically infers required inputs and outputs from the assembled blocks, providing a preview for user confirmation. This feature empowers users to create tailored workflows that meet their specific requirements.

The resource management view (Fig~\ref{fig:webui}.(d)) enables users to manage computational resources efficiently. As supported by the DST framework, users can connect to multiple servers, with the system automatically indexing the optimal server for executing algorithms. This ensures efficient resource utilization across distributed environments.

% The web WGUI provides a comprehensive, user-friendly interface for APL computations, catering to diverse user needs. By integrating advanced customization options and guided workflows, it bridges the gap between non-coding users and experienced researchers, making the APL pipeline accessible, flexible, and efficient.

\subsection*{Distributed RESTful API Framework}

% The Distributed RESTful API Framework (DRAF) is designed using the FastAPI framework~\cite{FastAPI}, enabling transformation of standard Python functions into RESTful API endpoints with auto-generated documentation. DRAF incorporates several essential features including:

\begin{figure}[!ht]%
    \centering
    \includegraphics[width=\linewidth]{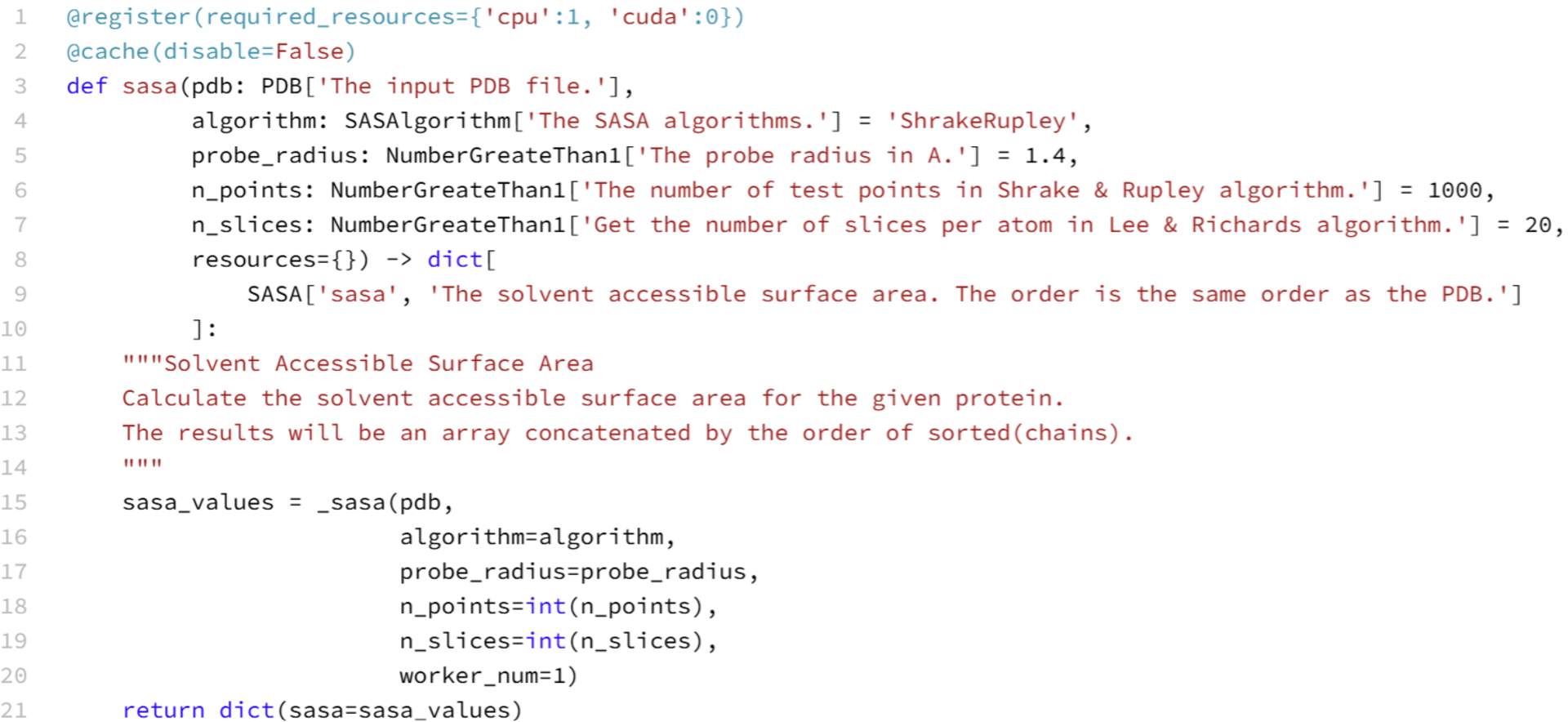}
    
    \caption{\textbf{Distributed RESTful API Framework (DRAF):} The Distributed RESTful API Framework (DRAF) enables seamless transformation of standard Python functions into RESTful API endpoints with auto-generated documentation.}
    \label{fig:draf}
\end{figure}

The Distributed RESTful API Framework (DRAF) as shown in Fig~\ref{fig:draf} is designed using the FastAPI framework~\cite{FastAPI}, enabling seamless transformation of standard Python functions into RESTful API endpoints with auto-generated documentation. DRAF incorporates several essential features to enhance efficiency, flexibility, and scalability:

\begin{itemize}
    \item Automatic conversion of Python functions into RESTful API endpoints.
    \item Automatic generation of endpoint documentation based on function parameters and descriptions.
    \item A parameter fingerprint-based caching mechanism to eliminate redundant computations.
    \item A resource-aware job scheduling system that optimizes the execution of computational tasks.
    \item A standardized data type protocol for consistent data exchange between APIs, servers, and clients, facilitating algorithm integration.
    \item Dual network communication support for standard HTTP requests and high-frequency WebSocket interactions.
\end{itemize}

\paragraph{Standardized Data Format for Algorithm Integration: }
DRAF enforces all data to follow a data protocol based on meta-types and rich types, to allow validating input and output data efficiently and chaining algorithms by connecting the output of one algorithm to the compatible input of another. The meta-types include numbers, strings, and numeric arrays, and the rich types are defined as subsets of these meta-types, with custom rules, descriptions, and names. For instance, Protein Data Bank (PDB) files can be defined as a subset of strings.
Moreover, it enables users to chain algorithms by connecting the output of one algorithm to the compatible input of another, facilitating the creation of complex, stacked workflows on the client side.

\paragraph{Automatic API Generation and Documentation: }
% DRAF leverages its standardized input and output definitions to automatically generate comprehensive documentation without requiring additional code. The meta-type standardization also facilitates parameter serialization, enabling the creation of unique hash fingerprints for parameter combinations. These fingerprints allow efficient caching of results, accelerating query responses while ensuring user privacy, as cached results are stored in a secure, non-readable format.
Unlike traditional RESTful API frameworks, DRAF leverages its standardized input and output definitions to simplify API creation. Developers can transform any Python function with typed parameters and return values into an API endpoint. Based on these type annotations and function comments, DRAF automatically generates comprehensive documentation without requiring additional code. The meta-type standardization also facilitates parameter serialization, enabling the creation of unique hash fingerprints for parameter combinations. These fingerprints allow efficient caching of results, accelerating query responses while ensuring user privacy, as cached results are stored in a secure, non-readable format.

\paragraph{Resource-Aware Job Scheduling: }
% Instead of executing tasks sequentially, DRAF organizes jobs into multiple queues, each optimized for specific computational resources to accommodate the diverse computational requirements of different algorithms. For example, a server equipped with two GPUs and ten CPUs could allocate one queue with two GPUs and five CPUs for GPU-intensive tasks (e.g., COREX), another queue with four CPUs for CPU-intensive tasks (e.g., BLAST), and a final queue with one CPU for lightweight tasks (e.g., SASA).
To accommodate the diverse computational requirements of different algorithms, DRAF employs a resource-dependent job scheduling system. Instead of executing tasks sequentially, DRAF organizes jobs into multiple queues, each optimized for specific computational resources. For example, a server equipped with two GPUs and ten CPUs could allocate one queue with two GPUs and five CPUs for GPU-intensive tasks (e.g., COREX), another queue with four CPUs for CPU-intensive tasks (e.g., BLAST), and a final queue with one CPU for lightweight tasks (e.g., SASA). This design minimizes runtime by assigning tasks to the most appropriate queue automatically.

\paragraph{WebSocket Support for Efficient Job Monitoring: }
% DRAF supports WebSocket communication to enable efficient periodic job status querying. WebSockets establish a persistent connection between the client and server, allowing the client to monitor job progress without sending repeated HTTP requests.
A common challenge with asynchronous API services is the need for clients to repeatedly query job statuses using HTTP requests, which can strain resources. To address this, DRAF supports WebSocket communication as an alternative. WebSockets establish a persistent connection between the client and server, allowing the client to monitor job progress without sending repeated HTTP requests. This reduces resource consumption and improves efficiency.

DRAF enables the transformation of Python functions with well-defined parameters into API endpoints that include automatic documentation, result caching, resource-aware job scheduling, and advanced communication protocols. Using DRAF, we have developed API endpoints for B-Factor, SASA, COREX, Sequence Entropy, BLAST, MHC-II prediction, and APL running on distributed servers.

\subsection*{Python Client and Data Science Tool}

\begin{figure}[!ht]%
\centering
\includegraphics[width=\linewidth]{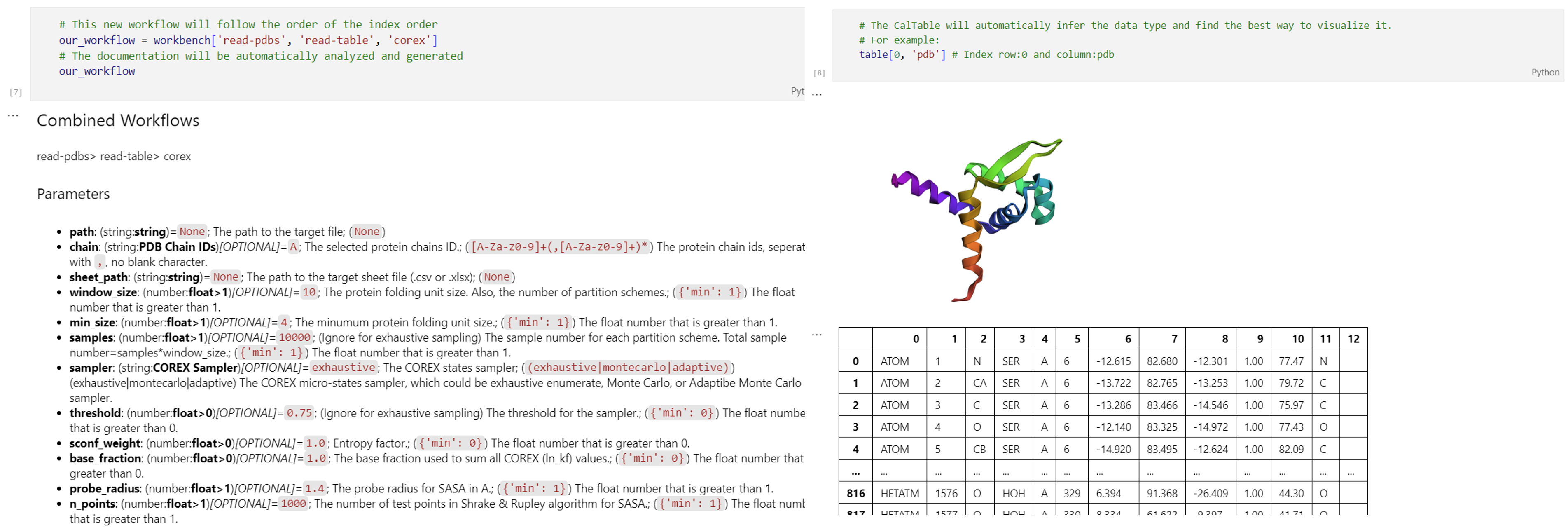}

\caption{\textbf{Data Science Tool (DST):} Data Science Tool (DST) streamlines access to DRAF’s functionality and make it more accessible to a wide range of users}
\label{fig:dst}
\end{figure}

% To fully leverage the capabilities of DRAF and enhance usability, we provide a Python client and a Data Science Tool (DST). The Python client acts as a bridge between users and DRAF by automatically exporting all API endpoints of a server as Python functions. Each function corresponds to an API endpoint, with documentation dynamically constructed by querying the server. This approach simplifies API usage, as users can interact with DRAF’s functionality directly within Python without dealing with HTTP requests or manual endpoint configuration. Built on top of the Python client, the Data Science Tool (DST) further enhances usability with multiple servers. Key features of the DST include:

To fully leverage the capabilities of DRAF and enhance usability, we provide a Python client and a Data Science Tool (DST) as shown in Fig~\ref{fig:dst}. These tools streamline access to DRAF’s functionality and make it more accessible to a wide range of users.

The Python client acts as a bridge between users and DRAF by automatically exporting all API endpoints of a server as Python functions. Each function corresponds to an API endpoint, with documentation dynamically constructed by querying the server. This approach simplifies API usage, as users can interact with DRAF’s functionality directly within Python without dealing with HTTP requests or manual endpoint configuration.

Built on top of the Python client, the Data Science Tool (DST) further enhances usability by enabling advanced operations and seamless integration with multiple servers. Key features of the DST include:

% \begin{itemize}
%     \item \textbf{Server Indexing and Load Balancing: } Users can connect to multiple DRAF servers, and the DST automatically chooses server for a remote API call.
%     \item \textbf{Algorithm Chaining: } Leveraging DRAF’s standardized data type protocol, the DST allows users to chain multiple algorithms to create complex workflow, and generates documentation for it based on the components' documentation. 
%     \item \textbf{Data Table Structure: } The DST introduces a Pandas-like data structure, called the data table, which enables users to perform operations, calculations, and visualizations on data.
%     \item \textbf{Basic Visualization: } Using meta-type definitions, the DST supports built-in visualizations for the data table. Users can view graphical representations of individual table cells without writing any additional code.
%     \item \textbf{Custom and Extensible Visualizations: } The DST supports user-defined visualizations and extensions, allowing users to tailor the tool to their specific needs. (e.g., 3D views of PDB protein structures)
% \end{itemize}
\begin{itemize}
    \item \textbf{Server Indexing and Load Balancing: } Users can connect to multiple DRAF servers, and the DST automatically identifies the best server for a remote API call. This functionality is as intuitive as indexing an element in a Python list.
    \item \textbf{Algorithm Chaining: } Leveraging DRAF’s standardized data type protocol, the DST allows users to chain multiple algorithms by providing a sequence of API endpoint indices. This simplifies the creation of complex workflows. In addition, for user customized chained algorithms, DSF will automatically generate a documentation based on the components' documentation. 
    \item \textbf{Data Table Structure: } The DST introduces a Pandas-like data structure, called the data table, which enables users to perform operations, calculations, and visualizations on data seamlessly. This familiar interface minimizes the learning curve for data manipulation.
    \item \textbf{Basic Visualization: } Using meta-type definitions, the DST supports built-in visualizations for the data table. Users can view graphical representations of individual table cells without writing any additional code. In addition, the protein related visualization extensions are embedded in the package for convenience.
    \item \textbf{Custom and Extensible Visualizations: } The DST supports user-defined visualizations and extensions, allowing users to tailor the tool to their specific needs. For example, rich visualizations such as 3D views of PDB protein structures or alignment views for FASTA sequences can be implemented or installed as extensions.
\end{itemize}

The Python client and DST work together to maximize the utility of DRAF by simplifying API access, enabling complex algorithmic workflows, and providing data manipulation and visualization capabilities. These tools empower users, ranging from novice coders to experienced researchers, to integrate and analyze computational results efficiently and effectively.

%% file: sections/apl.tex
\section*{Results}
To enhance the accessibility and utility of the APL computational pipelines, we integrated three categories of APL algorithms into the application: (1) fully automated APL pipelines, (2) APL pipelines incorporating MHC binding predictions (APL-MHC), and (3) standalone APL components. Both the APL and APL-MHC pipelines accept PDB files, PDB IDs, or AlphaFold-predicted mmCIF structures as inputs. These pipelines compute B-factors (or 100-pLDDT for AlphaFold mmCIF), solvent-accessible surface area (SASA), COREX stability, and sequence entropy. They subsequently generate residue- and peptide-level APL scores. The default parameters for APL aggregation is extracted from the CD4+ T cell immunodominance of influenza virus hemagglutinin research~\cite{cassotta2020deciphering}. The APL-MHC pipeline further integrates MHC binding predictions from the IEDB API based on user-specified alleles, combining these scores with APL results. As examples, the small proteins benchmark computed by APL and APL-MHC pipelines are provided with the application as shown in Fig~\ref{fig:webui}.e and Fig~\ref{fig:webui}.f.

To support modular usage, we provide independent tools for B-factor, 100-pLDDT, SASA, COREX, sequence entropy, residue APL, peptide APL, and MHC prediction. These tools are managed by a default local DRAF service, which is automatically deployed upon installation. Additionally, an independent DRAF server is available for installation on separate machines, enabling users to integrate it into their local workflows.

\subsection*{Single Antigen Analysis: Example with a Varicella Zoster antigen}
\begin{figure}[!ht]%
\centering
\includegraphics[width=0.48\linewidth]{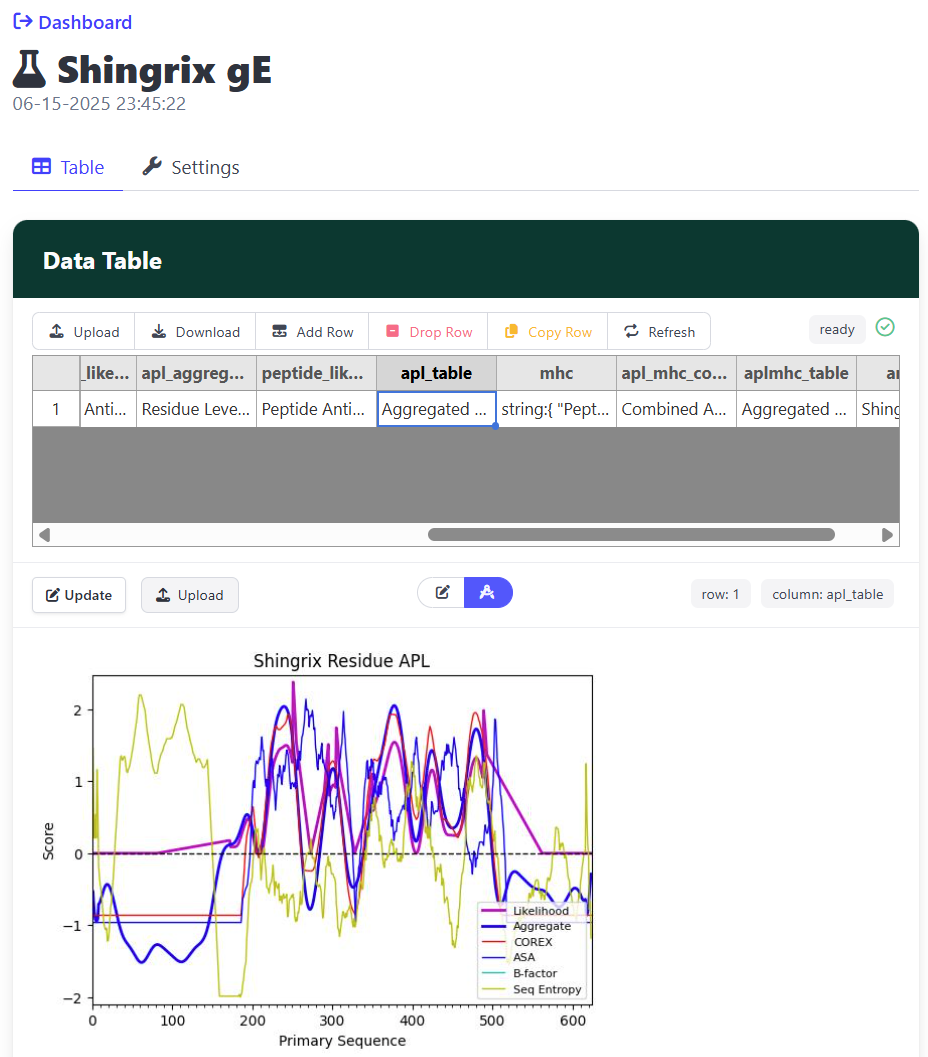}
\includegraphics[width=0.48\linewidth]{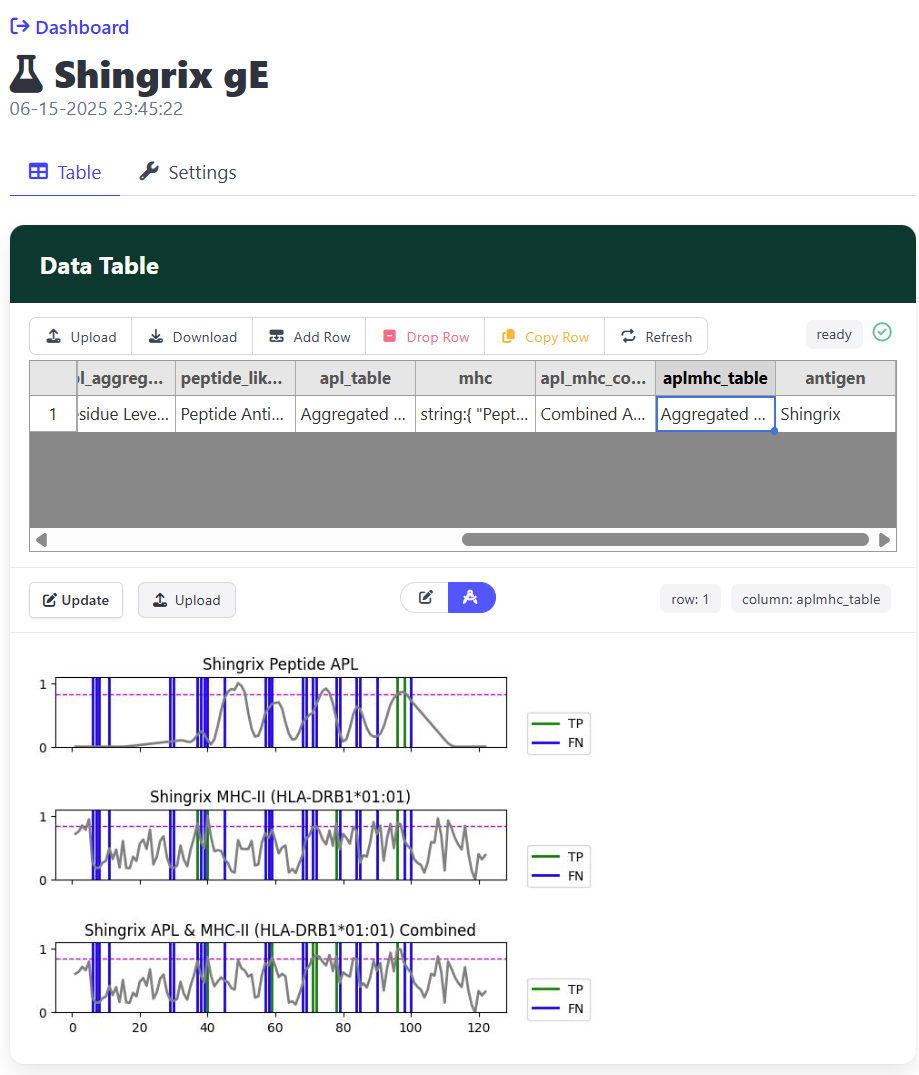}

\caption{\textbf{Application of the APL-MHC pipeline the VSV gE antigen.}}
\label{fig:shingrix}
\end{figure}
% Paraphrased from BIBM'23 
As an example of a typical use case for APLSuite analysis we consider the analysis of a single antigen: the glycoprotein E (gE) of Varicella Zoster Virus (VSV), a key antigen in the Shingrix vaccine~\cite{voic2020identification}. %Shingrix, a subunit vaccine targeting gE, has replaced Zostavax due to superior efficacy and applicability to immunocompromised individuals.
Figure~\ref{fig:shingrix} shows APL and MHC predictions along with mapped epitopes. 

We assessed APL, MHC-II binding (using seven high-performing alleles~\cite{paul2014development}), and a linear combination of both~\cite{bhattacharya2022incorporating}. This particular workflow is preloaded in the GUI and can be set up and dispatched in seconds for a single antigen. After processing, results can be saved in tabular and PDF formats for further analysis or presentation. Figure~\ref{fig:shingrix} shows results both at the residue and peptide levels, along with comparisons to ground truth epitopes (known in this case, but not required as input). 

%given antigen Among the top 12 predicted peptides, APL, MHC-II, and the combined model correctly identified 5, 3, and 8 epitopes, respectively. APL was most accurate in structured, stable regions (e.g., peptides 68–69, 96, 98, 100), while MHC-II captured epitopes in moderately unstable regions but also yielded false positives in highly unstable regions (e.g., peptides 3–4, 53). The combined method improved precision by leveraging both signal types, yielding the most accurate CD4+ epitope predictions.

\subsection*{Batch Processing: a Small Protein Benchmark}

\begin{figure}[!ht]%
\centering
\includegraphics[width=0.48\linewidth]{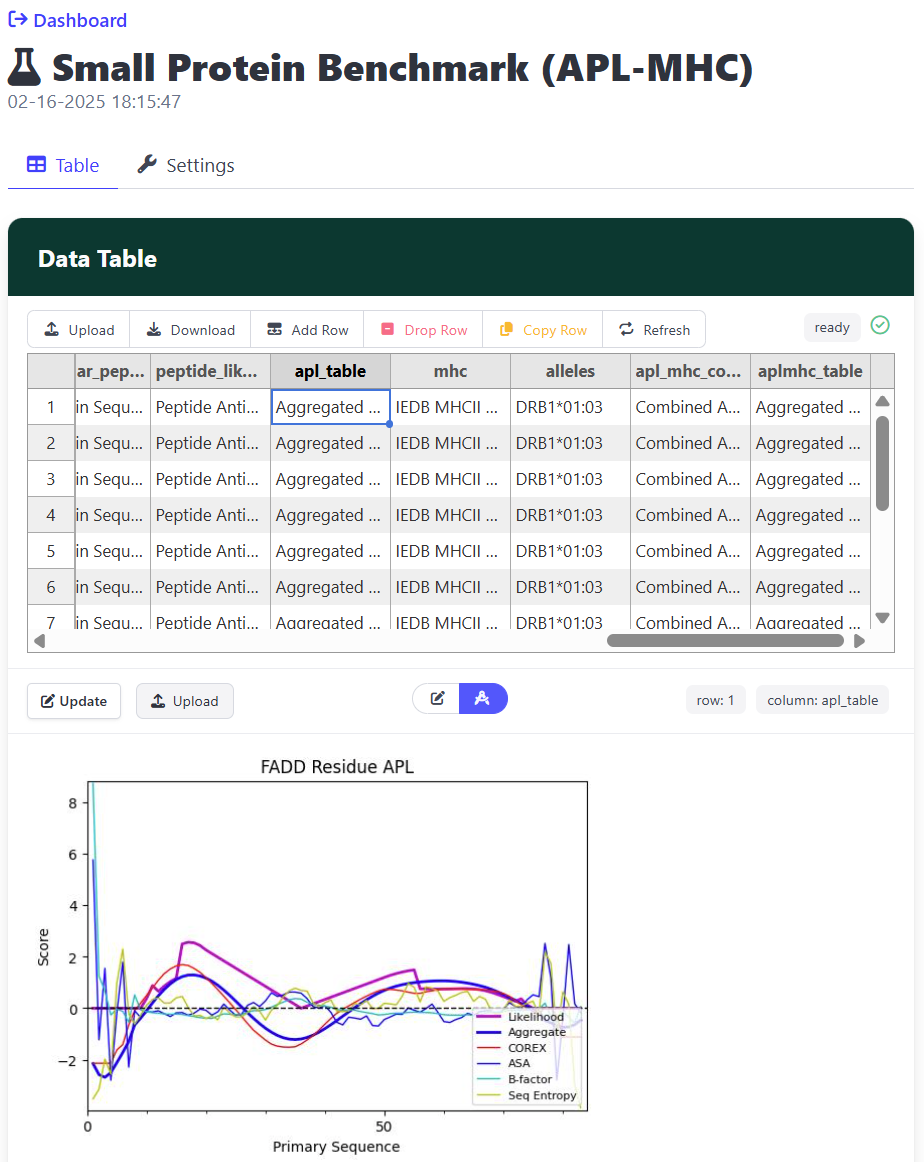}
\includegraphics[width=0.48\linewidth]{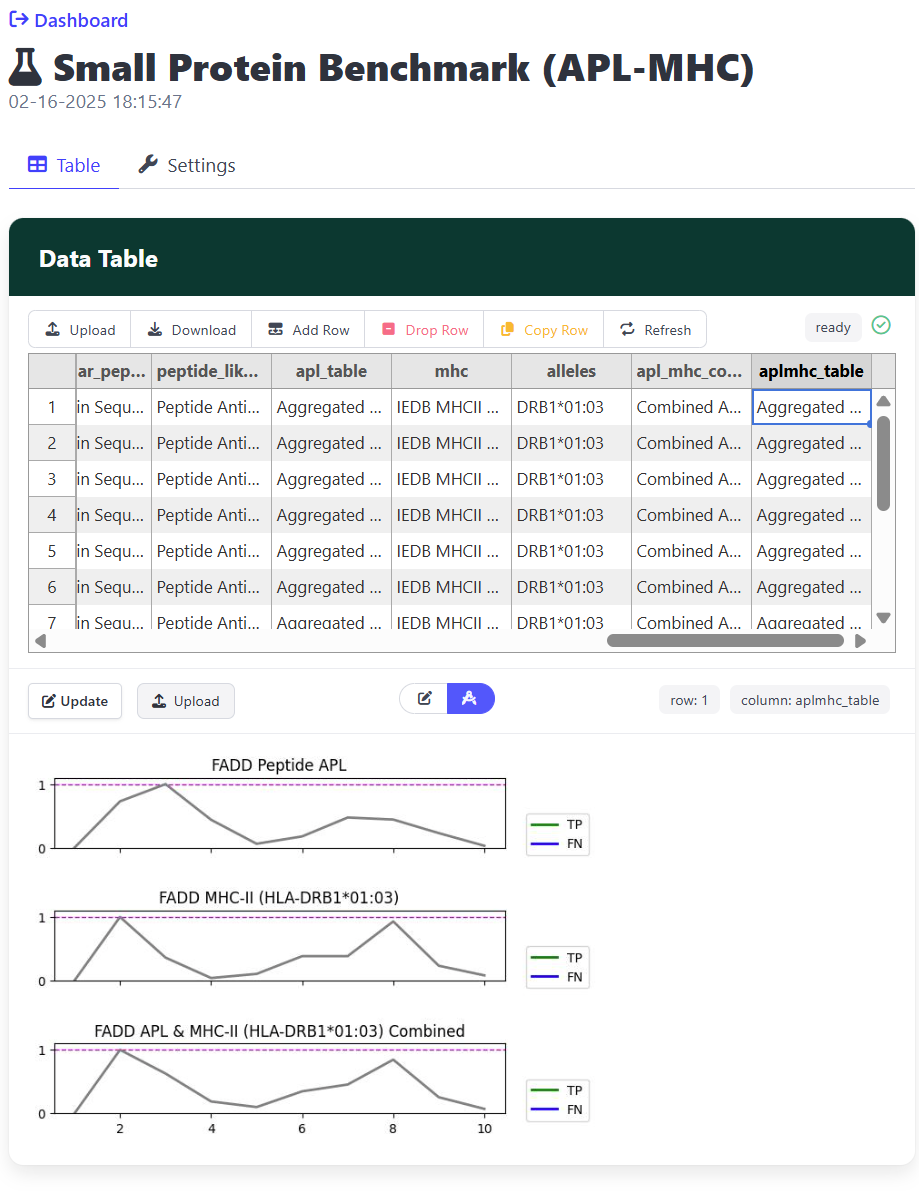}

\caption{\textbf{Application of the APL-MHC pipeline to a set of antigens.}}
\label{fig:spb}
\end{figure}

As part of the distribution we provide a  benchmark of small proteins (from~\cite{jiarui2024gpu}). This dataset compiled 13 small proteins ranging in size from 63 to 117 residues. Here we discuss how APLSuite can be used to conduct batch processing. The GUI provided with APLSuite enables the computation of APL and MHC binding of these proteins by setting up a single project that can be dispatched on numerous resources. Figure~\ref{fig:spb} shows a snapshot of the generated data table and results for one antigen. 

%% file: sections/conclusion.tex
\section*{Conclusion}
% APLSuite offers a series of tools including Distributed RESTful API framework (DRAF), Python client, data science tool (DST), and web application. It enables deploying different components of APL to various servers with different computational resources and transferring a Python function to API endpoints easily. The web application and data science tool enables table-like operation and automatically visualization. It makes non-coding users and novice coders to experienced researchers to utilize APL pipeline easily.

APLSuite provides an integrated suite of tools, including the Distributed RESTful API Framework (DRAF), a Python client, a Data Science Tool (DST), and a graphical user interface (GUI). This suite enables seamless deployment of APL components across multiple servers with diverse computational resources, transforming Python functions into API endpoints with minimal effort. The web application and DST offer intuitive, table-like operations and automatic data visualization, catering to users with varying levels of coding expertise. From non-coding users and novice programmers to experienced researchers, APLSuite simplifies access to and utilization of the APL pipeline, facilitating efficient computational workflows.

%% file: sections/availability.tex
\section*{Availability and Future Directions}
APLSuite is implemented in Python 3.12 with Django and FastAPI frameworks in the environment with BLAST, Clustal Omega, and Docker installed (or BLAST API). It is released under a \textit{GPLv3 license}. APLSuite can be downloaded from \url{https://github.com/Jiarui0923/APL} and run on any standard desktop computer or cloud services.

%% file: sections/acknowledgement.tex
\section*{Acknowledgments}
The authors thank the anonymous reviewers for their valuable suggestions.
We also acknowledge support from from Harold L. and Heather E.
Jurist Center of Excellence for Artificial Intelligence at Tulane University, and the Amazon AWS Cloud Credit for Research program for providing necessary GPU resources. 